\documentclass[12pt]{article}
\usepackage{amsfonts}
\usepackage{latexsym}
\usepackage{amsmath}
\usepackage{amssymb}

\newcommand{\newsection}{    % Numeration of eqs. is automatic
\setcounter{equation}{0}\section}
\def\appendix#1{\addtocounter{section}{1}\setcounter{equation}{0}
\renewcommand{\thesection}{\Alph{section}}
\section*{Appendix \thesection\protect\indent \parbox[t]{11.15cm}{#1}}
\addcontentsline{toc}{section}{Appendix \thesection\ \ \ #1}}

\newcommand{\be}{\begin{eqnarray}}
\newcommand{\ee}{\end{eqnarray}}
\newcommand{\bea}{\begin{eqnarray}}
\newcommand{\eea}{\end{eqnarray}}
\newcommand{\ba}{\begin{array}}
\newcommand{\ea}{\end{array}}
\newcommand{\nn}{\nonumber \\}

\newcommand{\la}{\label}

\def\a{\alpha}
\def\b{\beta}

\def\e{\epsilon}

\font\mybb=msbm10 at 11pt

\def\bb#1{\hbox{\mybb#1}}

\def\bZ {\bb{Z}}
\def\bR {\bb{R}}

\def\bH {\bb{H}}

\begin{document}
\begin{titlepage}
\begin{center}
%\today
\vspace{5.0cm}
%\hfill UB-ECM-PF-07-04
%\\
%\hfill hep-th/yymmnnn
%\\

\vspace{3.0cm} {\Large \bf (1,0) superconformal theories in six dimensions and  Killing spinor equations}
\\
[.2cm]

{}\vspace{2.0cm}
 {\large
M.~Akyol and  G.~Papadopoulos%$^2$,
 }

{}

\vspace{1.0cm}
%${}^2$
Department of Mathematics\\
King's College London\\
Strand\\
London WC2R 2LS, UK\\

\end{center}
{}
\vskip 3.0 cm
\begin{abstract}
We solve the Killing spinor equations of  6-dimensional (1,0) superconformal theories in all cases. In particular,
we derive the conditions on the fields imposed by the Killing spinor equations and demonstrate that these
depend on the isotropy group of the Killing spinors. We  focus on the models proposed
 by Samtleben et al in \cite{ssw} and find that there are solutions preserving 1,2, 4 and 8 supersymmetries.
 We also explore the solutions which preserve 4 supersymmetries
and find that many models admit string and 3-brane  solitons as expected from the M-brane intersection rules.
The string solitons are smooth regulated by the moduli  of  instanton configurations. 
\end{abstract}

\vfill
{{\small ~~~~ ~ mehmet.akyol@kcl.ac.uk}

{\small ~~~~ ~ george.papadopoulos@kcl.ac.uk}}

\end{titlepage}

\setcounter{section}{0}
\setcounter{subsection}{0}

%%%%%%%%%%%%%%%%%%%%%%%%%%%%%%%%%%%%%%%%%%%%%%%%%%%%%%%%%%%%%%%%%%%%%%%%%%

%%%%%%%%%%%%%%%%%%%%%%%%%%%%%%%%%%%%%%%%%%%%%%%%%%%%%%%%%%%%%%%%%%%%%%%%%%
\newsection{Introduction}

Recently there is some interest in understanding superconformal theories in 6 dimensions. This has been prompted  by the
 expectation that the worldvolume dynamics of multiple coincident  M5-branes is described by such a superconformal
theory. The main evidence for this is that  the near horizon geometry of the M5-brane supergravity solution \cite{gueven} is $AdS_7\times S^4$ \cite{gt} and so the theory which describes the worldvolume dynamics must exhibit a $SO(6,2)$ symmetry. Since $AdS_7\times S^4$ is a maximally supersymmetric solution of 11-dimensional supergravity, the  worldvolume theory must also have 16 supersymmetries or 32 fermionic symmetries after including
an additional 16 fermionic symmetries associated with the superconformal  generators.
In addition  for the multiple  M5-brane systems that are considered here,  the dynamics of the theory should be described
by  gauging  (2,0) tensor multiplet in 6 dimensions in analogy to similar constructions
for  M2-branes \cite{bl, gust}.  Such a (2,0) theory has been suggested in \cite{lp}, see  also \cite{chu2} for a bosonic theory. However, the closure of the supersymmetry algebra
requires that some of the fields must be independent from one of the worldvolume directions which makes the theory effectively 5-dimensional. In addition, as in the M2-brane case the construction
is based on Lorentzian 3-Lie algebras.  The description of dynamics in terms of Euclidean 3-algebras which leads to theories with better unitary
    behaviour is limited by the rigidity of their algebraic structure
   \cite{gp, gg} which does not allow for a general gauge group.
 As for M2-branes \cite{abjm}, one may consider M5-brane systems preserving less than maximal supersymmetry.
Following this,    Samtleben et al \cite{ssw} proposed a 6-dimensional (1,0)-supersymmetric
superconformal theory, see also \cite{chu}. The construction is based on gauging   (1,0)-tensor  multiplets and relies on the introduction of appropriate
 St\"uckelberg-type couplings. Some of the models constructed have a Lagrangian description provided one uses
a prescription to deal with the kinetic term of self-dual 3-form field strengths of (1,0) tensor multiplets.

 One of the results of this paper is the solution of  the Killing spinor equations (KSEs) of the (1,0) superconformal theories in 6-dimensions. In particular, we shall derive all the conditions on the fields in order a configuration preserves a fraction
of supersymmetry. The novelty of our approach relies on the systematic investigation of all possible conditions that can arise.
The supersymmetric configurations of a field theory include the solitons and instantons,  and so  are key ingredients  in the understanding
both its classical and quantum properties. For those theories that are associated to a brane construction, like for example those that describe the worldvolume dynamics of (multiple) branes, the  supersymmetric solutions have an additional interpretation in terms of  brane configurations  and so are instrumental in the understanding of the various phases of the theory.

We shall mostly focus on the models presented  in \cite{ssw}. However our method can be extended to all theories as it depends
on the properties of spinors in 6 dimensions instead of the details of the specific model. As a consequence the form of the KSEs is the same in all theories
  with 8 supercharges. Although the dependence of the various terms on the fields  may vary from theory to theory,  this does not affect the applicability of the spinorial geometry technique   \cite{uggp} as adapted to 6-dimensional  (1,0)  supergravity in \cite{ap1, ap2}
   which we have used
to solve the KSEs. In particular for the models in \cite{ssw}, we find that they  admit solutions preserving 1,2,4 and 8 supersymmetries. In all
cases, we explicitly  present the conditions imposed on the fields.

Next we shall focus on the solutions
 which preserve 4 supersymmetries, ie the half supersymmetric solutions. An inspection of the conditions that arise from the KSEs reveals that for all  6-dimensional (1,0) superconformal theories
  allow for the presence  of string and 3-brane solitons.  However, the existence of explicit solutions carrying appropriate charges is a model dependent property
 which depends on the couplings and the field content of a theory.  For the half
 supersymmetric solutions,  in addition to KSEs, we shall also investigate the
 field equations and Bianchi identities. In particular for the class of models in \cite{ssw}, we
 shall present explicit string and 3-brane soliton solutions. In particular, the string solutions are smooth and  supported by instantons.
 Clearly in an M-brane setting such worldvolume  solitons are expected on the grounds of M-brane intersection rules \cite{strominger,  gppt}. In particular,  M2-branes end on M5-branes on a self-dual
 string and two M5-branes intersect on a 3-brane.

This paper is organized as follows. In section 2, we present the field content, supersymmetry transformations
and KSEs of the models which we investigate later. In section 3, we give the solutions of the KSEs in all
cases. In section 4, we present the conditions on the fields for backgrounds preserving any number of supersymmetries.
In sections 5,6 and 7, we investigate the half supersymmetric solutions in a number of models and explicitly
give several string and 3-brane solutions. In section 8, we state our conclusions which include a brief discussion
on the applications to M-theory. In appendix A, we derive the field equations of the theory from the KSEs
using the Bianchi identities.

\newsection{(1,0) superconformal theory and KSEs}

\subsection{Fields and supersymmetry transformations}

The (1,0) superconformal model of \cite{ssw} has been constructed by gauging an arbitrary number of
tensor multiplets and the introduction of appropriate higher form fields which are used in Stuckelberg-type couplings.
The field content of the vector multiplets is $(A_\mu^r, \lambda^{ir}, Y^{ijr})$, where $r$ labels the different vector multiplets and $i, j = 1,2$ are the $Sp(1)$ R-symmetry indices, $A_\mu^r$ are 1-form gauge potentials, $\lambda^{ir}$ are symplectic Majorana-Weyl spinors and $Y^{ijr}$ are auxiliary fields.
The field content of the tensor multiplets is  $(\phi^I, \chi^{iI}, B_{\mu\nu}^I)$, where $I$ labels the different tensor multiplets, $\phi^I$ are scalars, $\chi^{iI}$ are symplectic Majorana-Weyl spinors, of opposite chirality from those
of the vector multiplets, and $B_{\mu\nu}^I$ are the 2-form gauge potentials.

The field strengths of the 1- and 2-form gauge potentials are
\be
\mathcal{F}_{\mu\nu}^r &\equiv& 2\partial_{[\mu}A_{\nu]}^r - f_{st}{}^rA_\mu^s A_\nu^t + h_I^rB_{\mu\nu}^I~,\\
\mathcal{H}_{\mu\nu\rho}^I &\equiv& 3D_{[\mu}B_{\nu\rho]}^I + 6d_{rs}^I A_{[\mu}^r\partial_\nu A_{\rho]}^s - 2f_{pq}{}^s d_{rs}^I A_{[\mu}^r A_\nu^p A_{\rho]}^q + g^{Ir}C_{\mu\nu\rho r}~,
\ee
where $f_{rs}{}^t$ are the structure constants,  $h_I^r, g^{Ir}$ and $d_{rs}^I=d_{(rs)}^I$ are St\"uckelberg-type couplings, and $C_{\mu\nu\rho r}$ are  three-form gauge potentials. The covariant derivative is defined as
\be
D_\mu \Lambda^s \equiv \partial_\mu \Lambda^s+ A_\mu^r(X_r)_t{}^s \Lambda^t~,~~~~D_\mu \Lambda^I \equiv \partial_\mu \Lambda^I+ A_\mu^r(X_r)_J{}^I \Lambda^J~,
\ee
where $X_r$ are given by
\bea
 (X_r)_t{}^s=-f_{rt}{}^s+d^I_{rt} h^s_I~,~~~(X_r)_J{}^I=2 h^s_J d^I_{rs}-g^{Is} b_{Jsr}~.
 \eea
 In addition,  covariance of the field strengths under the gauge transformations of gauge potentials requires that
 \bea
 2(d^J_{r(u} d^I_{v)s}-d^I_{rs} d^J_{uv}) h^s{}_J&=& 2 f_{r(u}{}^s d^I_{v)s}-b_{Jsr} d^J_{uv} g^{Is}~,
 \cr
 (d^J_{rs} b_{Iut}+ d^J_{rt} b_{Isu}+2 d^K_{ru} b_{Kst} \delta^J_I) h^u_J&=& f_{rs}{}^u b_{Iut}+f_{rt}{}^u b_{Isu}+ g^{Ju} b_{Iur}
 b_{Jst}~,
 \cr
 f_{[pq}{}^u f_{r]u}{}^s-{1\over3} h_I^s d^I_{u[p} f_{qr]}{}^u&=&0~,
 \cr
 h_I^r g^{Is}&=&0~,
 \cr
 f_{rs}{}^t h^r_I-d^J_{rs} h^t_J h^r_I&=&0~,
 \cr
 g^{Js} h_K^r b_{Isr}-2 h_I^s h_K^r d^J_{rs}&=&0~,
 \cr
 -f_{rt}{}^s g^{It} + d^J_{rt} h^s_J g^{It}- g^{It} g^{Js} b_{Jtr}&=&0~.
 \la{concon}
 \eea
It remains to give the supersymmetry transformations of the fields. Since, we are interested in the KSEs, it suffices to state
the supersymmetric variations of the fermions. These are given by
\be
%\delta A_\mu^r &=& -\bar{\epsilon}\Gamma_\mu \lambda^r~,\\
\delta \lambda^{ir} &=& \frac{1}{8}\mathcal{F}_{\mu\nu}^r\Gamma^{\mu\nu} \epsilon^i - \frac{1}{2}Y^{ijr}\epsilon_j + \frac{1}{4}h_I^r\phi^I\epsilon^i~,\\
%\delta Y^{ijr} &=& -\bar{\epsilon}^{(i}\Gamma^\mu D_\mu \lambda^{j)r} + 2h_I^r \bar{\epsilon}^{(i}\chi^{j)I}~,\\
%\delta \phi^I &=& \bar{\epsilon}\chi^I~,\\
\delta \chi^{iI} &=& \frac{1}{48}\mathcal{H}_{\mu\nu\rho}^I \Gamma^{\mu\nu\rho}\epsilon^i + \frac{1}{4}D_\mu\phi^I \Gamma^\mu \epsilon^i - \frac{1}{2}d_{rs}^I\Gamma^\mu \lambda^{ir}\bar{\epsilon}\Gamma_\mu \lambda^s~.
%\\
%\Delta B_{\mu\nu}^I &=& -\bar{\epsilon}\Gamma_{\mu\nu}\chi^I~,\\
%\Delta C_{\mu\nu\rho r} &=& -b_{Irs}\bar{\epsilon}\Gamma_{\mu\nu\rho}\lambda^s\phi^I~,
\ee
These as well as the remaining supersymmetry transformations  can be found in \cite{ssw}.
\subsection{Field equations }

The field equations of the minimal system are
\be
D^\mu D_\mu \phi^I &=& -\frac{1}{2}d_{rs}^I(\mathcal{F}_{\mu\nu}^r\mathcal{F}^{\mu\nu s} - 4Y_{ij}^rY^{ijs}) - 3d_{rs}^Ih_J^rh_K^s\phi^J\phi^K~,\label{eqa}\\
g^{Kr}b_{Irs}Y_{ij}^s\phi^I &=& 0~,\label{eqb}\\
g^{Kr}b_{Irs}\mathcal{F}_{\mu\nu}^s\phi^I &=&\frac{1}{4!}\epsilon_{\mu\nu\lambda\rho\sigma\tau}g^{Kr}\mathcal{H}_r^{(4)\lambda\rho\sigma\tau}~.
\label{eqc}
\ee
Observe that generically the theory has a cubic scalar field interaction and so the potential term is not bounded
from below. These field equations are also supplemented with
the Bianchi identities
\be
D_{[\mu}\mathcal{F}_{\nu\rho]}^r &=& \frac{1}{3}h_I^r\mathcal{H}_{\mu\nu\rho}^I~,\label{idf}\\
D_{[\mu}\mathcal{H}_{\nu\rho\sigma]}^I &=& \frac{3}{2}d_{rs}^I\mathcal{F}_{[\mu\nu}^r\mathcal{F}_{\rho\sigma]}^s + \frac{1}{4}g^{Ir}\mathcal{H}_{\mu\nu\rho\sigma r}^{(4)}~, \label{idh}
\ee
where $\mathcal{H}_{\mu\nu\rho\sigma r}^{(4)}$ is the field strength of the 3-form.

\subsection{KSEs}

The KSEs are the vanishing conditions for the supersymmetry variations of the fermions of the
theory evaluated at the locus where all fermions vanish. In this case, one finds that the KSEs are
\bea
\frac{1}{4}\mathcal{F}_{\mu\nu}^r\Gamma^{\mu\nu}\epsilon^i - Y^{ijr}\epsilon_j + \frac{1}{2}h_I^r\phi^I\epsilon^i &=& 0~,
\cr
\frac{1}{12}\mathcal{H}^I_{\mu\nu\rho}\Gamma^{\mu\nu\rho}\epsilon^i + D_\mu\phi^I\Gamma^\mu\epsilon^i &=& 0~.
\label{kse}
\eea
The first condition is  the vanishing condition of the supersymmetry variation of the fermions of the vector
multiplets while the second is  the vanishing condition of the supersymmetry variation of the fermions of the tensor
multiplets. In analogy with similar variations in 6-dimensional (1,0) supergravity, we shall refer to them as gaugini and
tensorini KSEs, respectively.

Before we proceed to solve these two KSEs, we note that all spinors that appear in the theory are symplectic Majorana-Weyl.
The gauge group of the theory is $Spin(5,1)\cdot Sp(1)$, where $Spin(5,1)=SL(2,\mathbb{H})$. In addition it is convenient
to rewrite the above KSEs using the formalism in \cite{ap1} where the symplectic Majorana-Weyl spinors have been identified with the $Sp(1)$-invariant $Spin(9,1)$ Majorana-Weyl spinors and realized as forms.
A basis for the
symplectic Majorana-Weyl spinors is
\bea
&&1+e_{1234}~,~~~i(1-e_{1234})~,~~~e_{12}- e_{34}~,~~~i(e_{12}+ e_{34})~,~~~
\cr
&&e_{15}+e_{2534}~,~~~i(e_{15}-e_{2534})~,~~~e_{25}-e_{1534}~,~~~i(e_{25}+e_{1534})~.~~~
\label{smw}
\eea
 We shall not give details of the construction for this see \cite{ap1}. Instead it suffices to introduce the $SU(2)$ generators
\be
\rho^1 = \frac{1}{2}(\Gamma_{38} + \Gamma_{49})~,~~~~~~\rho^2 = \frac{1}{2}(\Gamma_{89} - \Gamma_{34})~,~~~~~~\rho^3 = \frac{1}{2}(\Gamma_{39} - \Gamma_{48})~,
\label{rhos}
\ee
where the directions 3,4,8 and 9 are along an  auxiliary 4-dimensional space, and the 6 spacetime directions of the (1,0)
superconformal theory are $0,1,2,5,6$ and $7$.

In the above formalism the $i,j$ indices are suppressed and the term involving $Y$ in the KSEs can be re-written as
\be
-Y^{ijr}\epsilon_j = (Y^{r})_a \rho^a \epsilon^i~,
\ee
where $a=1,2,3$ span the $SU(2)$ generators, and we have made use of the fact that the $Sp(1)$ indices are raised and lowered by the antisymmetric tensors $\varepsilon_{ij}$ and $\varepsilon^{ij}$, with $\tau^i = \varepsilon^{ij}\tau_j$.
As a result the KSEs, (\ref{kse}), can be re-expressed as
\be
\frac{1}{4}\mathcal{F}_{\mu\nu}^r\Gamma^{\mu\nu}\epsilon + (Y^{r})_a\rho^a\epsilon + \frac{1}{2}h_I^r\phi^I\epsilon &=& 0~,\label{ksev12}\\
\frac{1}{12}\mathcal{H}^I_{\mu\nu\rho}\Gamma^{\mu\nu\rho}\epsilon + D_\mu\phi^I\Gamma^\mu\epsilon &=& 0~. \label{ksev22}
\ee
In what follows, we shall solve both KSEs for backgrounds preserving any number of supersymmetries.
\newsection{Killing spinors}

\subsection{Isotropy groups}

To identify the Killing spinors of the (1,0) superconformal theories, we shall first investigate the orbits of the
 gauge group $Spin(5,1)\cdot Sp(1)$ of the theory on the space of symplectic Majorana-Weyl spinors and identify the isotropy subgroups of spinors in
 $Spin(5,1)\cdot Sp(1)$.
This method has also been used in  \cite{ap1, ap2} to solve the KSEs of 6-dimensional (1,0)
supergravity. To find the invariant spinors and their isotropy groups,  observe that   the space of symplectic Majorana-Weyl spinors can be identified with $\mathbb{H}\oplus \mathbb{H}$. Then $Spin(5,1)=SL(2,\mathbb{H})$
acts on  $\mathbb{H}\oplus \mathbb{H}$ on the left with a $2\times 2$ quaternionic matrix multiplication while $Sp(1)$
acts with standard quaternionic multiplication on the right. Using this, it is straightforward to identify the
subgroups of $Spin(5,1)\cdot Sp(1)$ which preserve any number of spinors. The results have been tabulated in table 1.

\begin{table}[ht]
 \begin{center}
\begin{tabular}{|c|c|c|}
\hline
$N$&${\mathrm{Isotropy ~Groups}}$  & ${\mathrm{Invariant ~Spinors}}$ \\
\hline
\hline
$1$  & $Sp(1)\cdot Sp(1)\ltimes \bH$ & $1+e_{1234}$\\
\hline
$2$  & $(U(1)\cdot
Sp(1))\ltimes\bH$ & $1+e_{1234}~, ~i(1-e_{1234})$\\
\hline
$4$  & $
Sp(1)\ltimes \bH$ & $1+e_{1234}~, ~i(1-e_{1234})~,~e_{12}-e_{34}~,~i(e_{12}+e_{34})$\\
\hline
\hline
$2$  & $
Sp(1))$ & $1+e_{1234}~, ~e_{15}+e_{2345}$\\
\hline
$4$  & $
U(1)$ & $1+e_{1234}~,~i(1-e_{1234})~, ~e_{15}+e_{2345}~,~ i(e_{15}-e_{2345})$\\
\hline
\end{tabular}
\end{center}
%\label{tab1}
\label{ttt}
\caption{\small
The first column gives the number of invariant spinors, the second column the associated isotropy groups
and the third column representatives of the invariant spinors. Observe that if 3 spinors are invariant, then there is a fourth one
which is also invariant under the same isotropy group.
Moreover the isotropy group of more than 4 linearly independent spinors is the identity.}
\end{table}

\subsection{Killing spinor representatives}

\subsubsection{One Killing spinor}

It remains to identify the Killing spinors. If the KSEs (\ref{ksev12}) and (\ref{ksev22}) admit a Killing spinor, then
this can always be identified with $1+e_{1234}$. This is because $Spin(5,1)\cdot Sp(1)$ has one non-trivial orbit on
$\mathbb{H}\oplus \mathbb{H}$ with isotropy group $Sp(1)\cdot Sp(1)\ltimes \bH$. Next observe that the tensorini
KSE (\ref{ksev22}), if it admits one Killing spinor, then it admits 4. This is because it commutes with the
$\rho$ operations given in (\ref{rhos}). Moreover a basis for the 4 Killing spinors of (\ref{ksev22}) is given
by the $Sp(1)\ltimes\bH$ invariant spinors in table 1.

\subsubsection{Two Killing spinors}

To proceed, observe that $\mathbb{H}\oplus \mathbb{H}$ under the action of the isotropy group $Sp(1)\cdot Sp(1)\subset Sp(1)\cdot Sp(1)\ltimes \bH$ of the first Killing spinor decomposes as $\bR\oplus {\rm Im}\, \bH\oplus \bH$, where $\bR$ spans the first Killing spinor.
If the KSEs admit a second Killing spinor, then its linearly independent component must lie in ${\rm Im}\, \bH\oplus \bH$. In fact, this component must lie  either in ${\rm Im}\, \bH$ or in $ \bH$. This is because if it lies in both
${\rm Im}\, \bH$ and  $ \bH$, then one can use a $\bH\subset  Sp(1)\cdot Sp(1)\ltimes \bH$ gauge transformation to arrange
such that the component in ${\rm Im}\, \bH$ vanishes.
Now if the second Killing spinor lies in ${\rm Im}\, \bH$, then it can be arranged such that it is given
by $i(1-e_{1234})$. This is because $Sp(1)\cdot Sp(1)\subset Sp(1)\cdot Sp(1)\ltimes \bH$ acts on
${\rm Im}\, \bH$ with the 3-dimensional representation and so it can rotate any spinors to a particular direction.
Clearly a basis for the two Killing spinors can be identified with that of $U(1)\cdot Sp(1)\ltimes \bH$ invariant spinors
in table 1.

On the other hand the second Killing spinor may lie in $\bH$. In such a case, it can be chosen as
 $e_{15}+e_{2345}$. This is because $Sp(1)\cdot Sp(1)\subset Sp(1)\cdot Sp(1)\ltimes \bH$ acts on $\bH$ with left and right
 quaternionic multiplication and so any spinor can again be rotated to a particular direction. Therefore
 a basis of the two Killing spinors is given by that of $Sp(1)$ invariant spinors of table 1. In addition observe
 that in this case the tensorini KSE preserves all eight superasymmetries. This because it commutes with the
 $\rho$ operations (\ref{rhos}) but now acting on the two $Sp(1)$ invariant Killing spinors of table 1.

\subsubsection{Four Killing spinors}

Next suppose that the KSEs admit a third spinor and that the first two spinors are $U(1)\cdot Sp(1)\ltimes \bH$ invariant. Under $U(1)\cdot Sp(1)\subset U(1)\cdot Sp(1)\ltimes \bH$, $\bH\oplus \bH$ decomposes as $\bR\oplus \bR\oplus \bR^2\oplus \bH$, where the first two Killing spinors span $\bR\oplus \bR$. Therefore the linearly independent component
of the third spinor lies in $\bR^2\oplus \bH$. In fact, it can lie in either $\bR^2$ or $\bH$. This is because
if it lies in both, then there is a $\bH\subset U(1)\cdot Sp(1)\ltimes \bH$ transformation such that the component
in $\bR^2$ can be set to zero. If the third Killing spinor lies in $\bR^2$, then it can be chosen to lies in any direction
as $U(1)\subset U(1)\cdot Sp(1)\ltimes \bH$ acts with the standard 2-dimensional representation. In particular, one can choose as a third Killing spinor $e_{12}-e_{34}$. However, it can be shown that if  the gaugino KSE has solutions
$1+e_{1234}, i (1-e_{1234})$ and  $e_{12}-e_{34}$, then $i(e_{12}+e_{34})$ is also a solution. Since the tensorini
KSE also admits these four as Killing spinors, the number of supersymmetries preserved  are  enhanced from 3 to 4.

Next suppose that the first two Killing spinors are given by the $Sp(1)$ invariant spinors in table 1. Under $Sp(1)$,
$\bH\oplus \bH$ is decomposed as $\bR\oplus {\rm Im}\, \bH\oplus \bR\oplus {\rm Im}\, \bH$, where the first
two Killing spinors span the $\bR\oplus \bR$ subspace. Clearly, the linearly independent component of a third
Killing spinor lies in ${\rm Im}\, \bH\oplus {\rm Im}\, \bH$. As we have already mentioned, the tensorini KSE
admits 8 Killing spinors. Thus the investigation is focused on the gaugini KSE. For this, an analysis of the conditions that arise from the gaugini
 KSE on the two  $Sp(1)$ invariant spinors reveals that the two components of the third Killing spinors in
 ${\rm Im}\, \bH\oplus {\rm Im}\, \bH$ are independently Killing spinors. So without loss of generality, we can assume that
 the third Killing spinor lies in  the first  ${\rm Im}\, \bH$ of ${\rm Im}\, \bH\oplus {\rm Im}\, \bH$. Since
  $Sp(1)$ acts on each ${\rm Im}\, \bH$ with the 3-dimensional representation, the third Killing spinor can always be chosen to lie along  the $i(1-e_{1234})$ direction. If this is the case, one can show that the gaugino KSE admits $i(e_{15}-e_{2345})$
  as a Killing spinor and so there is an enhancement from 3 to 4 supersymmetries. A basis of the
  Killing spinors is given by that of the $U(1)$ invariant Killing spinors in table 1.

\subsubsection{More than four Killing spinors}

We shall not investigate this case in detail. However a straightforward analysis reveals that if a solution admits
more than 4 linearly independent Killing spinors, then it is maximally supersymmetric. The results of this section have been tabulated in table 2.

\begin{table}[ht]
 \begin{center}
\begin{tabular}{|c|c|c|}\hline
   ${\rm Isotropy ~Groups}$ &${\rm Gaugini} $& ${\rm Tensorini}$
 \\ \hline \hline
  $Sp(1)\cdot Sp(1)\ltimes\bH$& 1 &$4$\\
\hline
$U(1)\cdot Sp(1)\ltimes\bH$& 2 &$4$
\\ \hline
$Sp(1)\ltimes\bH$& 4&$4$
\\ \hline \hline
$Sp(1)$& 2&$8$
\\ \hline
$U(1)$& 4&$8$
\\ \hline
$\{1\}$&  8&$8$
\\ \hline
\end{tabular}
\end{center}
\caption{\small
In the first column the isotropy  groups of the Killing spinors of the gaugini KSE are given. In the second and third columns
  the number of Killing spinors of the gaugini and tensorini KSEs are stated, respectively. The isotropy groups of the Killing spinors of the tensorini KSE are either $Sp(1)\ltimes \bH$ or $\{1\}$.    The cases that do not appear in the table do not occur.}
\end{table}

\newsection{The solution of Killing spinor equations}

In this section, we shall derive the conditions imposed on the fields by the KSEs.
For this, it suffices to substitute into the KSEs the spinors given in table 1 and then solve
the resulting equations. This can be done in a straightforward way. In particular, one has the following.

\subsection{N=1 solutions}
The Killing spinor is $\epsilon = 1 + e_{1234}$. Substituting this  into gaugini KSE (\ref{ksev12}),  we find that the fields must satisfy
\be
\mathcal{F}_{-+}^r +  h_I^r\phi^I = 0~,~~~~~~~~~\mathcal{F}^r_{\alpha}{}^{\alpha} + 2i(Y^r)_1 = 0~,\nn
\mathcal{F}_{+\alpha}^r = 0 =\mathcal{F}_{+\bar{\alpha}}^r~,~~~~~~\mathcal{F}_{12}^r + (Y^{r})_2 - i(Y^r)_3 = 0~,
\label{1kse1}
\ee
where we have introduce lightcone and complex coordinates on Minkowski spacetime  $\bR^{5,1}$ as follows.  Recall that in the 10-dimensional description of the
spinors we have adopted the spacetime directions are along 0,5,1,6,2,7.
The Minkowski metric on $\bR^{5,1}$ has been  chosen as
\bea
ds^2&=&-(dx^0)^2+(dx^5)^2+ (dx^1)^2+(dx^6)^2+(dx^2)^2+(dx^7)^2
\cr
&=&2 e^- e^++ \delta_{ij} e^i e^j=2 e^- e^++2 \delta_{\a\bar\b} e^\a e^{\bar\b}~,~~~
\eea
where $e^\a$, $\a=1,2$, are the differentials of complex coordinates constructed from the pairs $(dx^1, dx^6)$ and $(dx^2, dx^7)$, respectively,
and $(e^-,  e^+)$ are the differentials of the lightcone coordinates along the directions $(dx^0, dx^5)$.
Observe that the $\mathcal{F}_{-i}^r$ components of the 2-form
field strength are not restricted by the KSEs. The same applies for the anti-self dual component  ${\cal F}^{\rm asd}$ of ${\cal F}_{ij}$. Moreover the
self-dual component of ${\cal F}$ is completely determined
in terms of the auxiliary field $Y$. Combining, the above results one can write
\bea
\mathcal{ F}^r&=&-h_I^r\phi^I \, e^-\wedge e^++\mathcal{ F}^r_{-i}\, e^-\wedge e^i- [(Y^{r})_2 - i(Y^r)_3] e^1\wedge e^2-[(Y^{r})_2 + i(Y^r)_3]e^{\bar 1}\wedge e^{\bar 2}
\cr
&+& (Y^r)_1 \omega+\mathcal{ F}^{{\rm asd},r}~,
\la{fum}
\eea
where $\omega=-i \delta_{\a\bar\b} e^\a\wedge e^{\bar\b}$.

Similarly, the tensorini KSE (\ref{ksev22}) gives
\be
D_{\bar{\alpha}}\phi^I + \frac{1}{2}\mathcal{H}^I_{-+\bar{\alpha}} + \frac{1}{2}\mathcal{H}^I_{\bar{\alpha}\beta}{}^{\beta} =0~,~~~
\mathcal{H}^I_{+\alpha\beta} = 0~,~~~\mathcal{H}^I_{+\alpha}{}^{\alpha} = 0~,~~~D_+\phi^I = 0~.
\label{1kse2}
\ee
The 3-form field strengths are restricted to be self-dual
\be
\mathcal{H}^I_{\mu\nu\rho} = \frac{1}{3!}\epsilon_{\lambda\sigma\tau\mu\nu\rho}\mathcal{H}^{I\lambda\sigma\tau}~.
\ee
Decomposing this condition in the $+,-, \a,\bar\a$ coordinates, one finds
\be
\mathcal{H}^I_{-+\alpha} + \mathcal{H}^I_{\alpha\beta}{}^{\beta} = 0~,~~~\mathcal{H}^I_{-+{\bar{\alpha}}} - \mathcal{H}^I_{\bar{\alpha}\beta}{}^{\beta} = 0~, ~\mathcal{ H}^I_{+1\bar 1}-\mathcal{ H}^I_{+2\bar2}=0~,~~~\mathcal{ H}^I_{+1\bar2}=0~.
\la{sdeqn}
\ee
Combining these conditions with those from the tensorini KSE, one finds that
\bea
\mathcal{H}^I_{+ij}=0~.
\eea
 $\mathcal{H}^I_{-ij}$ is {\it anti-self-dual} in the directions transverse to the light-cone and the remaining components
are determined in terms of $D\phi^I$. Therefore, one has
\bea
\mathcal{H}^I={1\over2} \mathcal{H}^I_{-ij}\, e^-\wedge e^i\wedge e^j- D_i\phi^I e^-\wedge e^+\wedge e^i+{1\over3!}
D_\ell\phi^I\,\epsilon^\ell{}_{ijk}\,\,e^i\wedge e^j\wedge e^k ~.
\la{hum}
\eea
Unlike the gaugini KSE, the tensorini KSE exhibits supersymmetry enhancement. In particular, if it admits
one Killing spinor $\epsilon$, it also admits three additional Killing spinors given by $\rho^1\e, \rho^2\e$ and $\rho^3\e$. For $\e=1+e_{1234}$, all four Killing spinors are given by the $Sp(1)\ltimes \bH$ invariant  spinors
of table 1.

\subsection{$N=2$ solutions with non-compact isotropy group}

The first Killing spinor is $\epsilon_1=1+e_{1234}$ as in the $N=1$ solutions described above. The second Killing spinor is $\epsilon_2=i (1-e_{1234})$ which
 can also be written as $\epsilon_2=\rho^1 \epsilon_1$, see table 1. Clearly for the solutions to admit two supersymmetries the KSEs must commute with
 the $\rho^1$ operation. As we have already mentioned, the tensorini KSE commutes with all the $\rho$ operations and so $\epsilon_2$ is
 also a Killing spinor. The 3-form field strength is given as in (\ref{hum}).

 This is not always the case for the gaugini KSE. In order for the gaugini KSE to commute with $\rho^1$,
 \bea
 (Y^{r})_2=(Y^r)_3 =0~.
 \la{y23}
 \eea
 The 2-form field strength is given as in (\ref{fum}) after imposing (\ref{y23}). Thus
 \bea
\mathcal{ F}^r=-h_I^r\phi^I \, e^-\wedge e^++\mathcal{ F}^r_{-i}\, e^-\wedge e^i+ Y^r \omega+\mathcal{ F}^{{\rm asd},r}~,
\la{fum2n}
\eea
 where we have set $Y^r=(Y^r)_1$.

\subsection{$N=2$ solutions with compact isotropy group }

The first Killing spinor is the same as that of $N=1$ solutions, $\epsilon_1=1+e_{1234}$. The second Killing spinor is $\epsilon_2=e_{15}+e_{2345}$.
We have already found the conditions imposed on the fields  by the KSEs evaluated on the first spinor. The conditions on the fields imposed
by the   (\ref{ksev12}) KSE evaluated on  $\epsilon_2$ are
\be
\mathcal{F}_{-+}^r -  h_I^r\phi^I = 0~,~~~~~~~~~\mathcal{F}^r_{1\bar{1}} - \mathcal{F}^r_{2\bar{2}} - 2i(Y^r)_1 = 0~,\nn
\mathcal{F}^r_{1\bar{2}} - (Y^r)_2 - i(Y^r)_3 = 0~,~~~~~~\mathcal{F}_{-\alpha}^r = 0~,~~~~~~\mathcal{F}_{-\bar{\alpha}}^r=0~.
\label{2ckse1}
\ee
Combining these conditions with those we have found for the Killing spinor $\epsilon_1$  in (\ref{1kse1}),  we get
\be
\mathcal{F}_{-+}^r=0~,~~~\mathcal{F}_{-i}^r=0~,~~~ \mathcal{F}_{+i}^r=0~,~~~ \mathcal{F}_{1\bar{1}}^r=0~,~~~ h_I^r\phi^I=0~,\nn
\mathcal{F}_{2\bar{2}}^r + 2i(Y^r)_1 = 0~,~~~\mathcal{F}_{12}^r + \mathcal{F}_{1\bar{2}}^r -2i(Y^r)_3 = 0~, ~~~\mathcal{F}_{12}^r - \mathcal{F}_{1\bar{2}}^r + 2(Y^r)_2 = 0~.
\ee
It is convenient to rewrite the above conditions in terms of real coordinates. In particular, we find
\be
\mathcal{F}_{-\nu}^r=0~,~~~ \mathcal{F}_{+\nu}^r=0~,~~~ \mathcal{F}_{\tilde{1}\nu}^r=0~,~~~ h_I^r\phi^I=0~,\nn
\mathcal{F}_{\tilde{6}\tilde{7}}^r = 2(Y^r)_2~,~~~\mathcal{F}_{\tilde{2}\tilde{6}}^r = 2(Y^r)_3 ~~~\mathcal{F}_{\tilde{2}\tilde{7}}^r = - 2(Y^r)_1~,
\la{fcon2}
\ee
where we have put a tilde on the real directions to distinguish them from the complex ones we have used so far to analyze the KSEs.
The above conditions on the fields can be most conveniently expressed by introducing a 3+3 split on the spacetime. In
 particular, we introduce coordinates $x^a$, $a=-,+, \tilde 1$ and $y^i$, $i=\tilde{2}, \tilde{6}, \tilde{7}$.  Using these,  (\ref{fcon2}) can be  written as
\be
\mathcal{F}_{ab}^r=0~,~~~ \mathcal{F}_{ai}^r=0~,~~~ h_I^r\phi^I=0~,~~~
\mathcal{F}_{ij}^r = -2\varepsilon_{ijk}(Y^r)^k~,
\ee
where $\varepsilon_{\tilde 2\tilde 6\tilde 7}= -1$ and we have set $(Y^r)^1=(Y^r)^{\tilde 6}$, $(Y^r)^2=(Y^r)^{\tilde 2}$ and $(Y^r)^3=(Y^r)^{\tilde 7}$.
Therefore we have
\bea
\mathcal{F}^r=-\varepsilon_{ijk}(Y^r)^k\, e^i\wedge e^j~,~~~h_I^r\phi^I=0~.
\la{fumn2c}
\eea

It remains to solve the tensorini KSE (\ref{ksev22}) evaluated on $\epsilon_2$. A straightforward calculation reveals that
\be
D_-\phi^I = 0~,~~~\mathcal{H}^I_{-1\bar{2}} = 0~,~~~\mathcal{H}^I_{-1\bar{1}}-\mathcal{H}^I_{-2\bar{2}} = 0~,~~~
\mathcal{H}^I_{-+\bar{\alpha}} - 2D_{\bar{\alpha}}\phi^I + \mathcal{H}^I_{\bar{\alpha}\beta}{}^\beta~ = 0~.
\ee
Combining these conditions with those we have found for $N=1$ solutions in (\ref{1kse2}) and the self-duality condition (\ref{sdeqn}),   we find that
\be
\mathcal{H}_{\mu\nu\rho}^{I} = 0~,~~~~~~~~~D_\mu\phi^I = 0~. \label{compact2conds}
\ee
Clearly in this case, the tensorini KSE preserves all 8 supersymmetries. Moreover, the integrability of the last condition
in (\ref{compact2conds}) implies that
\bea
F^r_{\mu\nu} X_{rJ}{}^I \phi^J=0~,
\eea
where $F^r_{\mu\nu}=2\partial_{[\mu}A_{\nu]}^r + X_{st}{}^r A_\mu^s A_\nu^t$.

\subsection{N=4 solutions with non-compact isotropy group }

The Killing spinors are the $Sp(1)\ltimes\bH$ invariant spinors of table 1. These can also be written as
\be
\epsilon_1 = 1 +e_{1234}~,~~~~~~~\epsilon_{2} = \rho^1\epsilon_1~,~~~\e_3=\rho^2 \e_1~,~~~\e_4=\rho^3\e_1~.
\ee
Clearly if $\epsilon_1$ is a Killing spinor, then the rest  are also Killing spinors provided
that the corresponding KSEs commute with the $\rho$ operations. As we have already mentioned this is the case
for the tensorini KSE (\ref{ksev22}) and the 3-form flux is given as in (\ref{hum}). Note also that $D_+\phi^I=0$.

The gaugini KSE commutes with all the $\rho$ operations iff all $Y$'s vanish, ie
\bea
Y^1=Y^2=Y^3=0~.
\eea
As a result using (\ref{fum}), the KSE implies that the 2-form flux is
\bea
\mathcal{ F}^r=-h_I^r\phi^I \, e^-\wedge e^++\mathcal{ F}^r_{-i}\, e^-\wedge e^i+ \mathcal{ F}^{{\rm asd},r}~.
\la{fn4nc}
\eea

\subsection{N=4 solutions with compact isotropy group}

The Killing spinors are the $U(1)$ invariant spinors of table 1.  Setting $\e_1=1+e_{1234}$ and $\e_2=e_{15}+e_{2345}$, the remaining two can be written as $\epsilon_3=\rho^1\epsilon_1$ and $\epsilon_4 = \rho^1\epsilon_2$. Thus
 these spinors solve the KSEs iff $\e_1$ and $\e_2$ are Killing spinors and the KSEs commute with the $\rho^1$
 operation.

 We have already shown that if $\e_1$ and $\e_2$ solve the tensorini KSE, then it preserves all supersymmetries.
 In particular both the 3-form flux and $D\phi$ vanish,
 \bea
 \mathcal{H}=D\phi=0~.
 \eea

 On the other hand the gaugini KSE commutes with $\rho^1$ iff $(Y^r)_2=(Y^r)_3=0$. Substituting this into
 (\ref{fumn2c}), we find that
 \be
\mathcal{F}^r=-2i Y^r e^2\wedge e^{\bar2}~,~~~~~~h_I^r\phi^I=0~,~~~~~~
\la{4ksecon}
\ee
where we have set $Y^r=(Y^r)_1$.

\subsection{Maximally supersymmetric solutions}
As we have mentioned all backgrounds which preserve more than 4 supersymmetries are maximally supersymmetric.
It is straightforward to see that the conditions on the fluxes for maximally supersymmetric backgrounds are
\bea
D_\mu\phi^I=0~,~~~h_I^r\phi^I = 0~,~~~\mathcal{F}_{\mu\nu}^r = 0~,~~~\mathcal{H}_{\mu\nu\rho}^{I} = 0~,~~~Y^{ijr} = 0~.
\eea
Thus all the scalars $\phi^I$ are  covariantly constant. In addition, those projected   by $h$ are required to vanish. Similarly the 2-form and 3-form
field strengths vanish as well. The same applies for the auxiliary fields $Y$.

\newsection{Half supersymmetric solutions without St\"uckelberg couplings}
\subsection{Summary of the conditions}

Before we proceed with the solution of the field equations and Bianchi identities for half supersymmetric backgrounds, we
shall first summarize the restrictions on the fields imposed by the KSEs. In particular, we have found that if the isotropy group
of the Killing spinors
is non-compact, then
\bea
\mathcal{F}^r&=&-h^r_I \phi^I e^-\wedge e^++\mathcal{F}_{-i} e^-\wedge e^i+\mathcal{F}^{{\rm asd},r}~,~~~D_+\phi^I=0~,
\cr
\mathcal{H}^I&=&{1\over2} \mathcal{H}_{-ij} e^-\wedge e^i\wedge e^j-D_i \phi^I e^-\wedge e^+\wedge e^i+{1\over 3!} D_\ell \e^\ell{}_{ijk}\, e^i\wedge
e^j\wedge e^k~,
\la{n4nc}
\eea
where all the auxiliary fields $Y$ vanish, $Y=0$,  and $\mathcal{H}_{-ij}$ and $\mathcal{F}^{{\rm asd},r}$ are anti-self-dual in the indices transverse to the light-cone.
On the other hand if the Killing spinors have compact isotropy group, then
\bea
\mathcal{H}^I&=&0~,~~~D_\mu\phi^I=0~,~~~h^r_I \phi^I=0~,
\cr
\mathcal{F}^r&=&-2i Y^r e^2\wedge e^{\bar 2}~,
\la{4nc}
\eea
where $(Y^2)^r=(Y^3)^r=0$ and $Y^r=(Y^1)^r$. In both cases above, we shall take the field equation (\ref{eqc}) as the definition
of $\mathcal{H}^{(4)}$.

\subsection{The model}

Let us first consider the model\footnote{This model does not have an apparent M-theoretic interpretation as it clearly
has dynamical vectors which are not thought to be present in a worldvolume theory of M5-branes. } with $g^{Ir}=h_I^r=0$. The conditions (\ref{concon}) are all satisfied provided
that $f$ are the structure constants of a Lie algebra $\mathfrak{g}$ and $d^I_{rs}$ and $b_{Irs}$ are invariant
symmetric tensors under the action of the adjoint representation of $\mathfrak{g}$. For example, one could set
$d^I_{rs}= d^I g_{rs}$ and $b_{Irs}=b_I g_{rs}$, where $g_{rs}$ is a bi-invariant metric on $\mathfrak{g}$.
Clearly in this case ${\cal F}$ is the standard curvature of a gauge connection.
To identify the half-supersymmetric solutions in this case, one has to solve both the Killing spinor and
field equations. There are two cases to consider depending on whether the isotropy group of the Killing spinors
is compact or not.

\subsection{Non-compact isotropy group}

The condition ${\cal F}_{+\mu}=0$ can be solved by fixing the gauge $A_+=0$ which then implies that $A_-,A_i$ do not
depend on the lightcone coordinate $x^+$. Similarly the scalars $\phi^I$ do not depend on $x^+$. Then the field equations
(\ref{eqa})  reduce to
\bea
\partial_i \partial^i \phi^I= -{1\over2} d^I_{rs} {\mathcal F}^r_{ij} {\mathcal F}^{ijs}~,
\la{scalarns}
\eea
where ${\mathcal F}^r_{ij}$ are anti-self-dual instantons along the directions transverse to the lightcone.
Observe that if  $g^{Ir}=h_I^r=0$, then the scalar fields are neutral (invariant) under the action of the gauge group and so they are not gauged.
If $d^I_{rs}= d^I g_{rs}$ and $b_{Irs}=b_I g_{rs}$, where $g_{rs}$ is a bi-invariant metric on $\mathfrak{g}$, then the right hand side  of (\ref{scalarns}) can be identified
with the Pontryagin density of instantons. In such a case, this equation can be solved for $\phi$ as the Pontryagin density of instantons with gauge groups like $SU(N)$ and $Sp(N)$
can be written as the Laplacian on a scalar\cite{osborn}. Similar equations have been solved before in the context of heterotic supergravity  \cite{halfhet} and
a more detailed analysis will be presented in section 7 which can be easily adapted to this case. Because of this,
we shall not further elaborate here on this.  The other two field equations (\ref{eqb}) and (\ref{eqc}) are automatically
satisfied.

It remains to solve the Bianchi identity (\ref{idh})  for $\mathcal{H}$ subject to the restrictions imposed by the
KSEs. The only independent component is
\bea
 \partial_-\partial_\ell\phi^I \epsilon^\ell{}_{ijk}-3 \partial_{[i}\mathcal{ H}^I_{jk]-}=6 d^I_{rs} \mathcal{F}^r_{-[i}  \mathcal{F}^s_{jk]}~.
 \la{bins}
\eea
This completes the analysis of the Bianchi identities.
\subsubsection{Strings and strings with waves solitons}

Suppose that a string spans the two light-cone directions. Such a class of solutions exhibits Poincar\'e invariance along the directions
  of the string. Such a class of solutions can be found by setting   $\mathcal{F}_{\pm\mu}=\mathcal{H}_{\pm\mu\nu}=0$  and requiring that all fields are independent from the
$x^\pm$ coordinates of the light-cone.
Clearly, the only non-trivial equation that remains to be solved is (\ref{scalarns}). As we have already mentioned this equations has solutions
provided that $d^I_{rs}= d^I g_{rs}$ and $b_{Irs}=b_I g_{rs}$ with gauge groups which include $SU(N)$ and $Sp(N)$ and for any instanton number.
Under certain conditions, all such solutions are regular. See section 7 for a more detailed analysis.

One can also find a more general solution by taking  $\mathcal{F}_{\pm\mu}=0$ and $\mathcal{H}_{-ij}\not=0$, and the fields $\mathcal{F}$ and $\phi$ to be independent
 of $x^\pm$. In such a case, the Lorentz invariance on the string is broken.
The component $\mathcal{ H}^I_{jk-}$ is restricted to be anti-self-dual and the Bianchi identity (\ref{bins}) requires that it should be closed.
So it can identified with an  abelian anti-self-dual field strengths on $\bR^4$ which are determined in terms of harmonic functions. There are no smooth solutions unless
 $\mathcal{ H}^I_{jk-}$ is taken to be constant. Such a solution has the interpretation of a string with a wave propagating on it.

 \subsection{Compact isotropy group}

 In this case, it follows from the tensorini KSE (\ref{4nc}) that the scalars $\phi^I$ are constant and $\mathcal{H}=0$. Moreover the auxiliary fields
 $Y^2=Y^3=0$ and the only non-vanishing component of $\mathcal {F}$ has support on a 2-dimensional
 subspace of the 4-dimensional space transverse to the light-cone directions. In this case, the KSE equations
 imply both the field equations and Bianchi identities. Therefore, the only non-trivial field is $\mathcal{F}_{2\bar2}$
 and it is related to $Y^1$ as in  (\ref{4ksecon}). Clearly, this solution exhibits a $\bR^{3,1}$ Poincar\'e invariance and so it has the
 interpretation of a 3-brane.

 \newsection{Half supersymmetric solutions of the adjoint model}

\subsection{The model}
The conditions in (\ref{concon}) are solved by taking the number of tensor multiplets to be the same
as the number of vector multiplets and setting
\bea
h^s_r=0~,~~~~d^s{}_{rt}= d_{prt} g^{ps}~,~~~b_{prt}=f_{prt}~,
\la{adcon}
\eea
where now $g$ is a bi-invariant metric, and $d$ is a totally symmetric bi-invariant tensor on the
Lie algebra $\mathfrak{g}$ with structure constants $f$. This model does not admit a Lagrangian description.

\subsection{Non-compact isotropy group}

Since the KSEs have been solved for the general model and the results have been summarized in  (\ref{n4nc}),  it remains to investigate the field equations and Bianchi identities of the model. Observe first that ${\mathcal F}$ is a standard curvature of the gauge connection. Since ${\mathcal F}_{+\mu}=0$, one can choose a gauge $A_+=0$ to find that all components of the gauge connection $A$ do not depend
on the lightcone coordinate $x^+$.

The field equation for the scalars is
\bea
D_i D^i \phi^r=-{1\over2} d^r{}_{st} \, {\mathcal F}^s_{ij} {\mathcal F}^{t,ij}~,
\eea
where we have used $\partial_+\phi^r=0$ which arises from the tensorini KSE.
Unlike the previous case, it is not apparent that anti-self-duality of ${\mathcal F}^s_{ij}$ implies
that the above equations has solutions. The equation depends on the second Casmir of  $\mathfrak{g}$ and so the analysis
requires the details of the Lie algebra of the gauge group used. This goes beyond the scope of this paper. The second
field equation (\ref{eqb}) is automatically satisfied as $Y=0$. One can view the last field
equation (\ref{eqc}) as the definition of $\mathcal{H}^{(4)}$. Substituting this into the Bianchi identity (\ref{idh})
and using the solution of the KSE in (\ref{n4nc}),
we find that the remaining independent equations are
\bea
 D_-D_\ell \phi^r \epsilon^\ell{}_{ijk}-3 D_{[i} \mathcal{H}^r_{jk]-}&=&6 d^r{}_{st}  \mathcal{F}^s_{-[i}
\mathcal{F}^t_{jk]}+\epsilon_{ijk}{}^m f^r{}_{st} \mathcal{F}^s_{-m}\phi^t
\cr
D_+ \mathcal{H}^r_{-ij}&=&0~,
\la{adn4}
\eea
where $\epsilon_{-+ijkl}=\epsilon_{ijkl}$ and $\epsilon_{-+1\bar1 2\bar2}=-1$. This concludes
the analysis of the Bianchi identities of the model.

\subsubsection{Strings and strings with waves solitons}

Solutions to (\ref{adn4}) can be easily found by
setting  $\mathcal{F}_{-\mu}=\mathcal{F}_{+\mu}=0$, choosing the gauge $A_\pm=0$,  identifying $H^r_{jk-}$
with the curvature of an anti-self-dual connection, and taking all fields to be independent from
the light-cone directions.

To find a solution of the theory, it remains to solve the field equations for the scalars. As it have been mentioned this
depends on the choice of gauge group. However, this can be circumvented in the special case where we choose the
coupling $d=0$. In such a case, the field equation for the scalars becomes
\bea
D_i D^i \phi^r=0~.
\la{linst}
\eea
A class of solutions of (\ref{linst}) is given by the Green functions of the Laplace operator in an instanton background.
These have been calculated for the adjoint representation in \cite{brown}, see also \cite{goddard}. However in such a case, the scalar equation
has delta function sources.

Alternatively, one can take the scalars in (\ref{linst})  to be neutral under the gauge group. This for example happens if
\bea
(X_r)_p{}^t=-g^{ts} b_{psr}~,
\la{hou}
\eea
vanishes on the active scalar fields of the solution.
Then the covariant Laplace equation above becomes a standard Laplace equation
and $\phi^r$ can be expressed in terms of harmonic functions. For the structure constants (\ref{hou}) on the active scalars to vanish,
 one may take $\mathfrak{g}=\mathfrak{t}\oplus \mathfrak{g}'$, where $\mathfrak{t}$ is an abelian algebra
which  commutes with the subalgebra $\mathfrak{g}'$, and the $\phi$'s and $\mathcal{H}$'s are restricted to take values in $\mathfrak{t}$.
Such solutions are singular unless $\phi$ is chosen to be constant.

Next if $H^r_{jk-}=0$,  the solutions above exhibit a $\bR^{1,1}$ Poincar\'e symmetry and so have an interpretation as strings.
On the other hand if $H^r_{jk-}\not=0$, the Poincar\'e symmetry is broken and the solutions have the interpretation
as waves propagating on strings.

\subsection{Compact isotropy group}

 The field equation for the scalars (\ref{eqa}) and the Bianchi identity (\ref{idf}) are satisfied
 either as a consequence of the conditions on the field imposed by the KSEs summarized in  (\ref{4nc}) or as a consequence of the restrictions  (\ref{adcon})
 on the coupling constants of the model.

 The Bianchi identity of the 3-form field strength (\ref{idh}) implies that $\mathcal{H}^{(4)}=0$
 as a consequence of the conditions imposed on the field by the  KSEs summarized in (\ref{4nc}).
 Then, the field equations  (\ref{eqb}) and (\ref{eqc}) require that
 \bea
 [\mathcal {F}, \phi]=0~.
 \eea
  As in the non-compact case, this condition  can be solved by taking $\mathfrak{g}=\mathfrak{t}\oplus \mathfrak{g}'$, where $\mathfrak{t}$
  commutes with the subalgebra $\mathfrak{g}'$, with the   scalars $\phi$  taking values in $\mathfrak{t}$ while $\mathcal{F}$
 takes values in $\mathfrak{g}'$.
   The only remaining condition is
 (\ref{4ksecon}) which relates $\mathcal {F}$ to the auxiliary field $Y$. Such solutions exhibit a $\bR^{3,1}$ Poincar\'e invariance and
 so have the interpretation of 3-branes.

\newsection{Half supersymmetric solutions of the $SO(5,5)$ model}

\subsection{The model}

We shall now investigate  models that admit a Lagrangian description\footnote{None of the theories which requires
a standard manifestly Lorentz invariant kinetic term for self-dual 3-form field strengths has a Lagrangian description. A Lagrangian
description is required for the rest of the fields modulo this well-known problem with the  3-form field strengths.}.  For this, there must exist
a metric $\eta_{IJ}$ such that
\bea
h^r_I=\eta_{IJ} g^{Jr}~,~~~d^I_{rs}={1\over2} \eta^{IJ} b_{Jrs}~.
\eea
The reduction of the conditions (\ref{concon}) to this case has been done in \cite{ssw} and it will not be repeated here.
In particular, we shall focus on the $SO(5,5)$ model of \cite{ssw} for which
\bea
b^I_{rs}=\gamma^I_{rs}~,~~~f_{rs}{}^t=-4 \gamma^{IJK}_{rs} \gamma_{IJp}{}^t g^p_K~,~~~g^{Ir} \gamma_{Irs}=0~,
\eea
where $\gamma^I_{rs}$ are the gamma-matrices of $SO(5,5)$, $\eta$ is the $SO(5,5)$-invariant Minkowski metric.
A key property of this model is that the cubic interaction of the scalars vanishes.

Before we proceed, we shall first describe some properties of the spinor representation of $SO(5,5)$ and
give an additional restriction on $g^{Ir}$ which is required in order for the coupling constants to solve (\ref{concon}).
A basis of the positive chirality $SO(5,5)$ spinors is
\bea
1,~~~e_{a_1a_2}~,~~~e_{a_1a_2a_3a_4}~,
\la{sbasis}
\eea
and the gamma matrices along the light-cone directions  act as
\bea
\gamma_a= \sqrt 2 e_a\wedge~,~~~\gamma_{\dot a}=\sqrt 2e_a\lrcorner~.
\eea
Therefore gamma matrices along the time-like and spacelike directions are
\bea
\Gamma_i=-e_i\wedge+ e_i\lrcorner~,~~~~\Gamma_{i+5}=e_i\wedge+ e_i\lrcorner~,~~~~i=1,2,3,4,5,
\eea
respectively, see \cite{uggp} for more details.
In this realization, the vector $SO(5,5)$ index decomposes as $I=(a,\dot a)$, and the Clifford
algebra relation is $\gamma_a\gamma_{\dot b}+\gamma_{\dot b} \gamma_a= 2\eta_{a \dot b}$. The Dirac inner
product $D (\psi, \chi):= <\Gamma_{12345}\psi, \chi>$  on the space of spinors gives
\bea
D(e_{a_1\dots a_k}, e_{a_{k+1}\dots a_5})=(-1)^{[{k+1\over 2}]+1} \epsilon_{a_1\dots a_5}~.
\eea
Observe that the inner product is skew-symmetric in the interchange of pairs.
This can be used to raise and lower spinor indices as
\bea
\psi_{b_1\dots b_{5-k}}&:=&\psi^{a_1\dots a_k} D_{a_1\dots a_k, b_1\dots b_{5-k}}= {(-1)^{[{k+1\over 2}]+1}\over k!}\psi^{a_1\dots a_k} \epsilon_{a_1\dots a_k b_1\dots b_{5-k}}
\cr
&:=&\eta_{b_1 \dot b_1}\dots
\eta_{b_{5-k}\dot b_{5-k}} \psi^{\dot b_1\dots \dot b_{5-k}}~.
\eea
In this realization, the positive chirality spinor representation decomposes as ${\bf 1}+ {\bf 10}+{\bf 5}$ under
the subgroup $SO(5)\subset GL(5, \bR)\subset SO(5,5)$ acting on the light-cone decomposition of the
vector representation of $SO(5,5)$ presented above.  The restriction on $g^{Ir}$ is that its only non vanishing
component lies in the ${\bf 15}$ representation, ie the non-vanishing component is
\bea
g^{\dot a \dot b}=-{1\over 4!} \epsilon^{\dot b}{}_{b_1\dots b_4} g^{\dot a b_1\dots b_4}~,~~~g^{\dot a \dot b}=g^{(\dot a \dot b)}
\eea
Clearly $\eta_{IJ} g^{Ir}  g^{Js}=0$ as it is required by the condition (\ref{concon}) on the couplings.

\subsection{Non-compact isotropy group}

Taking that the cubic scalar interaction term vanishes and using the conditions on the fields
imposed by the KSEs described in  (\ref{n4nc}), one finds that the field equation for the scalars
can be rewritten as
\bea
D_i D^i\phi^I=-{1\over2} d^I_{rs} \mathcal{F}^r_{ij} \mathcal{F}^{s,ij}~.
\eea
To expand this in $SO(5)$ representations observe that
\bea
(\gamma_a)_{b_1\dots b_k, b_{k+1}\dots b_4}&=& (-1)^{[{k\over 2}]}\sqrt 2 \, \epsilon_{a b_1\dots b_4}~,~~~k=0, \dots 4~,
\cr
(\gamma_{\dot a})_{b_1\dots b_k, b_{k+1}\dots b_6} &=& (-1)^{[{k\over2}]+1} k \sqrt{2} \delta_{\dot a[b_1}
 \epsilon_{b_2\dots b_k]b_{k+1}\dots b_6}~,~~~k=1, \dots, 5~.
\eea
and  that the gamma
matrices are symmetric in the interchange of spinor indices. In addition, a spinor is expanded in the basis (\ref{sbasis})
as
\bea
\Psi=  \psi\, 1 + {1\over2} \psi^{ab} e_{ab}+{1\over 4!} \psi^{a_1\dots a_4} e_{a_1\dots a_4}~.
\eea

Using these,  the field equation for the scalars can be rewritten as
\bea
D_i D^i\phi^{\dot a}&=&{\sqrt2 \over2} {\mathcal F}_{ij}\, {\mathcal F}^{\dot a, ij}+ {\sqrt2 \over16}
\e^{\dot a}{}_{b_1b_2b_3b_4} {\mathcal F}^{b_1b_2}_{ij}\,  {\mathcal F}^{b_3b_4, ij}~,
\cr
D_i D^i\phi^{a}&=&{\sqrt2 \over2} {\mathcal F}^{ab}{}_{ij}\,  {\mathcal F}_{b}{}^{ij}~.
\eea
The second field equation  (\ref{eqb}) follows from the KSEs as the latter imply that all auxiliary fields $Y$ vanish.
The third field equation, (\ref{eqc}), can be seen as the definition of $\mathcal{H}^{(4)}$. In particular,
one has
\bea
g^{\dot a\dot b}\mathcal{H}^{(4)}_{\mu_1\dots \mu_4, \dot b}=-{\sqrt2 \over2} \e_{\mu_1\dots \mu_4}{}^{\nu_1\nu_2}[- g^{\dot a\dot b} \phi_{\dot b} \mathcal{F}_{\nu_1\nu_2}+
g^{\dot a}{}_{b_1} \phi_{b_2} \mathcal{F}^{b_1b_2}{}_{\nu_1\nu_2}]~.
\eea
Next using the expression for $\mathcal{ F}$ in  (\ref{n4nc}), we find
\bea
g^{\dot a\dot b}\mathcal{H}^{(4)}_{ijk\ell, \dot b}=g^{\dot a\dot b}\mathcal{H}^{(4)}_{+ijk, \dot b}=0~.
\eea
Moreover the rest of the components are determined in terms of $\phi$ and $\mathcal{F}$.

Let us now turn to the investigation of Bianchi identities (\ref{idf}) and (\ref{idh}). In particular, the former implies that
\bea
D_{[\mu_1} \mathcal{F}_{\mu_2\mu_3]}=D_{[\mu_1}\mathcal{ F}^{ab}_{\mu_2\mu_3]}=0~,~~~D_{[\mu_1} \mathcal{F}^{\dot a}_{\mu_2\mu_3]}={1\over3} g^{\dot a}{}_b\mathcal{H}^b_{\mu_1\mu_2\mu_3}~.
\la{fbianchi}
\eea
The first two conditions give in particular
\bea
D_+\mathcal{F}_{-i}=D_+\mathcal{F}_{ij}=0~,~~~D_+\mathcal{F}^{ab}_{-i}=D_+\mathcal{F}^{ab}_{ij}=0~.
\eea
Similarly using (\ref{hum}), the last equation in (\ref{fbianchi}) gives
\bea
D_+\mathcal{F}^{\dot a}_{-i}=D_+\mathcal{F}^{\dot a}_{ij}=0~,
\eea
and
\bea
D_-\mathcal{F}^{\dot a}_{ij}+2 D_{[i}\mathcal{F}^{\dot a}_{j]-}=g^{\dot a \dot b} \mathcal{H}_{-ij, \dot b}~,~~~3\,D_{[i}\mathcal{F}^{\dot a}_{jk]}=g^{\dot a\dot b}\, D_\ell \phi_{\dot b} \, \e^\ell{}_{ijk}~.
\eea

Next let us turn to the Bianchi identity for $\mathcal {H}$ (\ref{idh}). This decomposes as
\bea
D_{[\mu_1} \mathcal{H}^a_{\mu_2\mu_3\mu_4]}&=&{3\over4} \gamma^a_{rs} \mathcal {F}^r_{[\mu_1\mu_2} \mathcal{F}^s_{\mu_3\mu_4]}
\cr
D_{[\mu_1} \mathcal{H}^{\dot a}_{\mu_2\mu_3\mu_4]}&=&{3\over4} \gamma^{\dot a}_{rs} \mathcal {F}^r_{[\mu_1\mu_2} {F}^s_{\mu_3\mu_4]} +{1\over 4 \cdot 4!} g^{\dot a b_1\dots b_4}
\mathcal {H}^{(4)}_{\mu_1\dots\mu_4, b_1\dots b_4}
\eea

The independent conditions which arise from the above Bianchi identities are
\bea
D_+\mathcal{H}^a_{ijk}=0~,~~~D_+\mathcal{H}^a_{-jk}=0~,~~~D_-\mathcal{H}^a_{ijk}-3 D_{[i} \mathcal{H}^a_{jk]-}=3 \gamma^a_{rs} \mathcal {F}^r_{-[i} \mathcal {F}^s_{jk]}
\eea
and
\bea
D_+\mathcal{H}^{\dot a}_{ijk}&=&0~,~~~D_+\mathcal{H}^{\dot a}_{-jk}=0~,~~~
\cr
D_-\mathcal{H}^{\dot a}_{ijk}-3 D_{[i} \mathcal{H}^{\dot a}_{jk]-}&=&
-3\sqrt 2 [\mathcal {F}_{-[i} \mathcal {F}^{\dot a}_{jk]} +\mathcal {F}^{\dot a}_{-[i} \mathcal {F}_{jk]} + {1\over4} \e^{\dot a}{}_{b_1b_2c_1c_2}
\mathcal {F}^{b_1b_2}_{-[i} \mathcal {F}^{c_1c_2}_{jk]}]
\cr
&-& \sqrt2\e_{ijk}{}^\ell [g^{\dot a}{}_b \phi^b \mathcal{F}_{-\ell}+g^{\dot a}{}_{b_2} \phi_{b_1}
\mathcal{F}^{b_1b_2}_{-\ell}]~.
\eea
This concludes the analysis of the field equations and Bianchi identities of the theory.

\subsubsection{Regular string solutions}

This system has  a string solution. Suppose that the string lies along the two light-cone directions, and take that
\bea
{\cal F}={\cal F}^{\dot a}=0~, ~~~{\cal F}^{ab}_{\pm\mu}=0~.
\eea
The latter condition is required because of Lorentz invariance along the worldvolume directions of the string.

Moreover, choose the gauge $A^r_\pm=0$ and assume all non-vanishing fields to be independent of the light-cone coordinates $x^\pm$.
Using these, the field equations and the Bianchi identities above reduce to
\bea
D_i D^i \phi^{\dot a}={\sqrt 2\over 16} \e^{\dot a}{}_{b_1\dots b_4} \mathcal{F}_{ij}^{b_1b_2} \mathcal{F}^{b_3b_4,ij}~,~~~D_i D^i \phi^{ a}=0~,
\cr
g^{\dot a}{}_b D_i \phi^b=0~,~~~g^{\dot a\dot b} \mathcal{H}_{-ij, \dot b}=0~,~~~D_{[i}\mathcal{H}^I_{jk]-}=0~.~~~
\eea
To proceed, one moreover demands that
\bea
D_i \phi^b=0~,~~~\mathcal{H}^I_{jk-}=0~.
\eea
The latter condition is again required by Lorentz invariance along the string.
The integrability condition of the first condition requires that
\bea
\mathcal{F}_{ij}^{ab} g_{bc} \phi^c=0~.
\eea
One solution is to take $\phi^a$ constant with $g_{bc} \phi^c=0$. For simplicity, we shall take $\phi^a=0$.
Then the only remaining non-trivial equation is
\bea
D_i D^i \phi^{\dot a}={\sqrt 2\over 16} \e^{\dot a}{}_{b_1\dots b_4} \mathcal{F}_{ij}^{b_1b_2} \mathcal{F}^{b_3b_4,ij}~,
\la{soscalar}
\eea
where $\mathcal{F}_{ij}^{b_1b_2}$ is an anti-self-dual connection with gauge group $SO(5)$.

To solve (\ref{soscalar})
 choose $\phi^{\dot a}$ to lie along the 5-th direction and $\mathcal{F}_{ij}^{b_1b_2}$ to have gauge group
$SO(4)\subset SO(5)$ orthogonal to $\phi^{\dot 5}$. Now there are two cases to consider. First, if one restricts $\mathcal{F}_{ij}^{b_1b_2}$ to lie in one
of the $\mathfrak{su}(2)$ subalgebras of $\mathfrak{so}(4)$, $\mathfrak{so}(4)=\mathfrak{su}(2)\oplus\mathfrak{su}(2)$, then the
field equation (\ref{soscalar}) can be rewritten as
\bea
\partial_i \partial^i \phi^{\dot 5}=\pm{\sqrt 2\over 8}  \mathcal{F}_{ij,b_1b_2} \mathcal{F}^{b_1b_2,ij}~,~~~b_1, b_2=1,2,3,4~,
\la{fffeq}
\eea
where the sign depends on the choice of $\mathfrak{su}(2)$ subalgebra. Such an equation has been solved before in the context of NS5-branes
in heterotic supergravity wrapped on $AdS_3\times S^3$ in \cite{halfhet} and relies on the
 well-known property of the Pontryagin density of an $SU(2)$ anti-self-dual connection to be written as the Laplace operator
on a function. In particular choosing the minus sign and writing
\bea
\mathcal{F}^{b_1b_2}_{ij}=J^{b_1b_2}_{r'} \mathcal{F}_{ij}^{r'}~,~~~r'=1,2,3~,
\eea
where $J_{r'}$ is a basis of constant anti-self-dual 2-forms in $\bR^4$, the equation (\ref{fffeq}) can be rewritten as
\bea
\partial_i \partial^i \phi^{\dot 5}=-{\sqrt 2\over 2} \delta_{r's'}\mathcal{F}^{r'}_{ij} \mathcal{F}^{s',ij}~.
\eea
This equation can be solved for all $SU(2)$ instantons which follows from the well-known property of the Pontryangin
density to be written as the Laplace operator on a scalar \cite{osborn}. The non-vanishing fields of the
solution are
\bea
&&\mathcal{F}^{ab}={1\over 2} \mathcal{F}^{ab}_{ij} e^i\wedge e^j~,~~
\mathcal{H}^{\dot 5}=- \partial_i\phi^{\dot 5} e^-\wedge e^+\wedge e^i+{1\over3!}
\partial_\ell\phi^{\dot 5}\,\epsilon^\ell{}_{ijk}\,\,e^i\wedge e^j\wedge e^k ~,~~
\cr
&&\phi^{\dot 5}=\phi^{\dot 5}(x)~,~~~a,b=1,2,3,4~.
\eea
Observe that $g^{Ir} \mathcal{H}_r^{(4)}=0$.

To present an explicit solution, we consider the  configuration  with instanton number 1 and use the results of \cite{halfhet}. In such a case, the gauge connection $A$ of $\mathcal{F}^{ab}$ and $\phi^{\dot 5}$
can be written as
\bea
A^{ab}&=&2 (J^{r'})^{ab} (J_{r'})_{ij} {x^j\over |x|^2+\rho^2}\,\, e^i~,~~~\phi^{\dot 5}=\phi_0+ 4\sqrt 2{|x|^2+2\rho^2\over (|x|^2+\rho^2)^2}+ h_0~,~~~
\cr
h_0&=&\sum_\nu{Q_\nu\over |x-x_\nu|^2}~,
\eea
where $x$ are the coordinates in $\bR^{5,1}$ transverse to the light-cone where the string lies, $\phi_0$ is a constant, and $\rho$ is the instanton modulus. Moreover   $h_0$ is a multi-centred harmonic function, which if it is included in the solution, then delta function sources have to be added in the field equation for $\phi$.
Let us focus on the solution with $h_0=0$.  Such a solution is smooth. At large $|x|$, ie far away from the string, the scalar $\phi$ becomes constant, and the gauge connection pure gauge. As $|x|$ becomes small,
the values of $\phi$ and $A$ are regulated by the modulus $\rho\not=0$ of the instanton. Similarly, all our solutions with any instanton number $k$ are smooth. Compare this with the solution of \cite{howe} which is singular at the position of the string defect. 

The dyonic string charge $q_s$ of all solutions can be computed by integrating $\mathcal{H}^{\dot 5}$ on the 3-sphere at infinity. After an appropriate normalization\footnote{See \cite{halfhet} for a comment on the normalization of string charges in a similar setting.}, this can be
identified with the instanton number $k$, ie
\bea
q_s=\int_{S^3\subset \bR^4} \mathcal{H}^{\dot 5}=k~.
\eea

A more general solution can be found by taking $\mathcal{F}$ to take values in both $\mathfrak{su}(2)$ subalgebras of $\mathfrak{so}(4)$.
In such a case, the scalar field equation can now be rewritten as
\bea
\partial_i \partial^i \phi^{\dot 5}=-{\sqrt 2\over 2} \delta_{r's'}(\mathcal{F}^{r'}_{ij} \mathcal{F}^{s',ij}-\tilde{\mathcal{F}}^{r'}_{ij} \tilde{\mathcal{F}}^{s',ij})~,
\eea
where $\tilde{\mathcal{F}}$ are the anti-self-dual gauge fields associated with the second $\mathfrak{su}(2)$.

The 1-instanton solutions are now modified as
\bea
A^{ab}&=&2 (J^{r'})^{ab} (J_{r'})_{ij} {x^j\over |x|^2+\rho^2}\,\, e^i+ 2(I^{r'})^{ab} (J_{r'})_{ij} {x^j\over |x|^2+\sigma^2}\,\, e^i   ~,~~~
\cr
\phi^{\dot 5}&=&\phi_0+ 4\sqrt 2{|x|^2+2\rho^2\over (|x|^2+\rho^2)^2}-4\sqrt 2{|x|^2+2\sigma^2\over (|x|^2+\sigma^2)^2}+ h_0~,~~~
\cr
h_0&=&\sum_\nu{Q_\nu\over |x-x_\nu|^2}~,
\eea
where $I^{r'}$ is a basis in the space of self-dual 2-forms in $\bR^4$ and $\sigma$ is the modulus of the $\tilde{\mathcal{F}}$ instanton. The rest of the
notation is the same as in the previous solution. If we set $h_0=0$, the solution is again smooth. The same applies for all such solutions constructed
from $SU(2)$ instanton with any instanton number.

The dyonic string charge of this solution can also be computed using a similar calculation as in the previous solution. In particular, one finds that
\bea
q_s=\int_{S^3\subset \bR^4} \mathcal{H}^{\dot 5}=k-\tilde k~,
\eea
where $k$ and $\tilde k$ are the instanton numbers of ${\mathcal{F}}$ and $\tilde{\mathcal{F}}$, respectively. If in the original model only the
$SO(4)\subset SO(5)$ subgroup is gauged, then $q_s$ is a gauge invariant charge of the theory and so it is expected to be conserved in dynamical
processes. It is amusing that there are non-trivial solutions\footnote{Even though they are supersymmetric
some further analysis is required to show that they are stable.} with zero overall string charge.

\subsection{Compact isotropy group}

Taking as in the non-compact case that the cubic scalar interaction term vanishes, the  conditions implied by the
KSEs on the fields in  (\ref{4nc}) imply the field equation
for the scalars (\ref{eqa}). Similarly,   the Bianchi identity for $\mathcal{H}$ (\ref{idh}) implies  that
\bea
g^{Kr} \mathcal{H}_r^{(4)}=0~,
\eea
and the Bianchi identity for $\mathcal{F}$  (\ref{idf}) is  automatically satisfied. The remanning conditions on the fields implied
by the KSEs and  field equations
are
\bea
D_\mu\phi^I=0~,~~~g^{Kr}b_{Irs}Y^s\phi^I =0~,~~~h^r{}_K\phi^K=0~.
\la{lll}
\eea
Note that $\mathcal{H}=0$.
The integrability of the first equation requires that
\bea
(F \phi)^I=0~,
\la{iii}
\eea
ie the scalars must be invariant under the holonomy  group of the gauge connection.

So far the analysis has been independent of the particular model and applies to all solutions
which preserve 4 supersymmetries.
In the particular model we are examining, (\ref{lll}) can be investigated further. First, (\ref{iii}) can be rewritten as
\bea
\mathcal{F}^{ab} (X_{ab})_J{}^I \phi^J=0~.
\la{incon}
\eea
For this equation  to have solutions either $\phi=0$ or the holonomy group of the gauge connection $A^{ab}$ reduces to a subgroup of  $SO(5)$.
 If $\phi=0$, then
 \bea
 \phi^I=\mathcal{H}^I=0~,~~~g^{Kr} \mathcal{H}_r^{(4)}=0~,~~~\mathcal{F}^r= -2iY^r\,  e^2\wedge e^{\bar 2}~,
 \eea
 is a solution for an arbitrary auxiliary field $Y$ which depends on the complex coordinates $(x^2, x^{\bar 2})$.

 Alternatively, suppose that the holonomy of $A^{ab}$ reduces to $SO(4)$. In that case, the constant scalar field $\phi=(\phi^5, \phi^{\dot 5})$ solves the integrability condition (\ref{incon}) and the first condition in (\ref{lll}).  Furthermore the
 last condition in (\ref{lll}) implies that $g^{\dot a}{}_5 \phi^5=0$ and so we take $\phi^5=0$. The second condition
 in (\ref{lll}) is automatically satisfied. Thus
 \bea
 \phi=(0, \phi^{\dot 5})~,~~~\mathcal{F}^r= -2iY^r\,  e^2\wedge e^{\bar 2}~,~~~{\rm hol}(\mathcal{F}^{ab})\subseteq SO(4)~,
 \eea
 is a solution for  auxiliary fields $Y$ which satisfy the holonomy condition and depend on the complex coordinates $(x^2, x^{\bar 2})$.
 Other solutions can be constructed by taking further reductions of the holonomy group of the gauge connection.
 The above solutions are invariant under
the $\bR^{3,1}$ Poincar\'e group and so they have the naive interpretation of 3-branes.

\newsection{Concluding remarks}

We have solved  the KSEs of all (1,0) superconformal theories in 6 dimensions and presented the conditions imposed
on the fields by supersymmetry  in all cases. We have mostly focused on the models presented in \cite{ssw} but our results apply more generally.
We have found that such theories admit solutions which preserve 1,2,4 and 8 supersymmetries.
We have investigated in detail the conditions on the fields which arise for half supersymmetric solutions and we have presented several
 explicit solutions which include strings and 3-branes.
The models investigated so far do not include hyper-multiplet couplings but one can extend  the analysis
to include those using the classification of supersymmetric solutions of 6-dimensional (1,0) supergravities in \cite{ap1}.

It is not apparent that the models presented in \cite{ssw} describe the dynamics of multiple M5-branes. First,
they have 8 supersymmetry charges than 16 which are required for an effective theory of multiple M5-branes
propagating in flat space. However, this may be circumvented by considering the dynamics of M5-branes on a suitable
orbifold which breaks half of the supersymmetry. For example, one can place the M5-brane on a $\bZ_2$ orbifold.  In addition, the models found so far which admit a Lagrangian description, modulo the usual problem with the self-dual 3-forms, do not have the correct number of scalars to describe the transverse
directions of M5-branes in flat space or an orbifold. This is because the only dynamical  scalars  are  those 
 of the tensor multiplets. For the description of all 5 transverse directions,  one has to include one hypermultiplet
  for each tensor multiplet. Nevertheless, the models explored so far can describe a multiple M5-brane system 
  where four transverse scalars are fixed at the singular point of an orbifold, ie the vacuum of a potential,  and so the only
  modulus scalars are those of the tensor multiplets. Such an interpretation has some attractive features. In particular, we have shown that the models admit
string solutions which are in accordance with one of the M-brane intersection rules.  This states that
a M2-brane ends on a M5-brane on a string \cite{strominger} which appears as a defect of the M5-brane worldvolume theory. Moreover, our string solutions are smooth regulated by the moduli
of instantons, ie the smoothness of the solutions is directly related to the non-abelian properties of the multiple M5-brane system.  
The models we have investigated  also admit 3-brane solutions which again are in accordance with the intersection rule that two M5-branes
intersect on a 3-brane \cite{gppt}. The 3-brane appears as a defect of the M5-brane effective theory. However for a better
understanding of the 3-brane solitons in relation to M-brane intersection rules, one has to include also  hyper-multiplets
as one requires the presence of two active transverse scalars per M5-brane. These two scalars are associated with the
worldvolume directions of the incoming M5-brane.

\vskip 1.0cm
\noindent{\bf Acknowledgements} \vskip 0.1cm
\noindent
We thank Neil Lambert and Henning Samtleben for helpful discussions. Part of this work was done during the participation of one of us, GP, at the Newton Institute's   ``The  mathematics and applications of branes in string and M-theory'' programme.
MA is supported by a STFC studentship.
GP is partially supported by the STFC rolling grant ST/G000/395/1.
\vskip 0.5cm

\setcounter{section}{0}
\setcounter{subsection}{0}

\appendix{The integrability conditions of the KSEs}
The KSEs are given by
\be
\frac{1}{4}\mathcal{F}_{\mu\nu}^r\Gamma^{\mu\nu}\epsilon + (Y^{r})_a\rho^a\epsilon + \frac{1}{2}h_I^r\phi^I\epsilon &=& 0~,\label{ksev12a}\\
\frac{1}{12}\mathcal{H}^I_{\mu\nu\rho}\Gamma^{\mu\nu\rho}\epsilon + D_\mu\phi^I\Gamma^\mu\epsilon &=& 0~. \label{ksev22a}
\ee
The field equations of the minimal system can be written as
\be
D^\mu D_\mu \phi^I &=& -\frac{1}{2}d_{rs}^I(\mathcal{F}_{\mu\nu}^r\mathcal{F}^{\mu\nu s} - 8Y_a^rY^{s,a}) - 3d_{rs}^Ih_J^rh_K^s\phi^J\phi^K~,\\
g^{Kr}b_{Irs}Y_{ij}^s\phi^I &=& 0~,\\
g^{Kr}b_{Irs}\mathcal{F}_{\mu\nu}^s\phi^I &=&\frac{1}{4!}\epsilon_{\mu\nu\lambda\rho\sigma\tau}g^{Kr}\mathcal{H}_r^{(4)\lambda\rho\sigma\tau}~.
\ee

We will now show how these field equations can be obtained from the KSEs and the Bianchi identities. We first square the KSE in (\ref{ksev12a}) as follows
\be
\left(\frac{1}{4}\mathcal{F}_{\mu\nu}^r\Gamma^{\mu\nu} - (Y^{r})_a\rho^a + \frac{1}{2}h_I^r\phi^I\right)\left(\frac{1}{4}\mathcal{F}_{\rho\sigma}^s\Gamma^{\rho\sigma}\epsilon + (Y^{s})_b\rho^b\epsilon + \frac{1}{2}h_J^s\phi^J\epsilon\right) &=& 0~,
\ee
multiplying through with $d_{rs}^I$ and simplifying we find
\be
\frac{1}{4}d_{rs}^I\mathcal{F}_{\mu\nu}^r\mathcal{F}_{\rho\sigma}^s\Gamma^{\mu\nu\rho\sigma}\epsilon - \frac{1}{2}d_{rs}^I\mathcal{F}_{\mu\nu}^r\mathcal{F}^{s,\mu\nu}\epsilon +4d_{rs}^I Y^r_aY^{s,a}\epsilon &+&\nn d_{rs}^I\mathcal{F}_{\mu\nu}^rh_J^s\phi^J\Gamma^{\mu\nu}\epsilon + d_{rs}^Ih_J^rh_K^s\phi^J\phi^K\epsilon &=& 0~.
\ee
Furthermore, we make use of the duality that the gamma matrices satisfy in six dimensions
\be
\Gamma^{\mu_1...\mu_n}\epsilon = \frac{(-1)^{[{n\over2}]+1}}{(6-n)!}\epsilon^{\mu_1...\mu_n}{}_{\nu_1...\nu_{6-n}}\Gamma^{\nu_1...\nu_{6-n}}\epsilon~,
\ee
using this for the case of $n=4$ the equation above becomes
\be
-\frac{1}{8}d_{rs}^I\mathcal{F}_{\mu\nu}^r\mathcal{F}_{\rho\sigma}^s\epsilon^{\mu\nu\rho\sigma}{}_{\lambda\tau}\Gamma^{\lambda\tau}\epsilon - \frac{1}{2}d_{rs}^I\mathcal{F}_{\mu\nu}^r\mathcal{F}^{s,\mu\nu}\epsilon +4d_{rs}^I Y^r_aY^{s,a}\epsilon &+&\nn d_{rs}^I\mathcal{F}_{\mu\nu}^rh_J^s\phi^J\Gamma^{\mu\nu}\epsilon + d_{rs}^Ih_J^rh_K^s\phi^J\phi^K\epsilon &=& 0~.
\label{kse1sq}
\ee

Now we act on the KSE in (\ref{ksev22a}) with $\Gamma^\mu D_\mu$ and this gives
\be
\frac{1}{12}D_\mu \mathcal{H}^I_{\nu\rho\sigma}\Gamma^{\mu\nu\rho\sigma}\epsilon + \frac{1}{4}D^\mu \mathcal{H}^I_{\mu\nu\rho}\Gamma^{\nu\rho}\epsilon + D_\mu D_\nu \phi^I\Gamma^{\mu\nu}\epsilon + D_\mu D^\mu \phi^I \epsilon = 0~.
\label{kse2cond}
\ee
The third term in this equation can be written as
\be
D_\mu D_\nu \phi^I\Gamma^{\mu\nu}\epsilon = d_{rs}^I\mathcal{F}^r_{\mu\nu}h_J^s\phi^J\Gamma^{\mu\nu}\epsilon - \frac{1}{2}\mathcal{F}^r_{\mu\nu}g^{Is}b_{Jsr}\phi^J\Gamma^{\mu\nu}\epsilon~.
\label{term3}
\ee
In addition the first and second terms in (\ref{kse2cond}) can be rewritten using the duality of the gamma matrices and the self-duality of the 3-form field strength.
Combining the results of this with (\ref{term3}) means the equation in (\ref{kse2cond}) becomes
\be
-\frac{1}{12}D_\mu \mathcal{H}^I_{\nu\rho\sigma}\epsilon^{\mu\nu\rho\sigma}{}_{\lambda\tau}\Gamma^{\lambda\tau}\epsilon + d_{rs}^I\mathcal{F}^r_{\mu\nu}h_J^s\phi^J\Gamma^{\mu\nu}\epsilon-\frac{1}{2}\mathcal{F}^r_{\mu\nu}g^{Is}b_{Jsr}\phi^J\Gamma^{\mu\nu}\epsilon + D_\mu D^\mu \phi^I \epsilon = 0~.
\label{cond2}
\ee
Subtracting this equation from (\ref{kse1sq}) we get
\be
\left[ D_\mu D^\mu \phi^I + \frac{1}{2}d_{rs}^I\mathcal{F}_{\mu\nu}^r\mathcal{F}^{s,\mu\nu} - 4d_{rs}^I Y^r_aY^{s,a} - d_{rs}^Ih_J^rh_K^s\phi^J\phi^K \right]\epsilon&&~\nn
-\frac{1}{2}\mathcal{F}^r_{\mu\nu}g^{Is}b_{Jsr}\phi^J\Gamma^{\mu\nu}\epsilon &&~\nn
-\frac{1}{12}\left[ D_\mu \mathcal{H}^I_{\nu\rho\sigma} - \frac{3}{2}d_{rs}^I\mathcal{F}_{\mu\nu}^r\mathcal{F}_{\rho\sigma}^s\right] \epsilon^{\mu\nu\rho\sigma}{}_{\lambda\tau}\Gamma^{\lambda\tau}\epsilon&=&0~.
\label{int12}
\ee
To proceed we make use of another identity that is obtained when the gaugini KSE (\ref{ksev12a}) is contracted with $g^{Ir}b_{Jrs}\phi^J$,
\be
g^{Ir}b_{Jrs}\mathcal{F}^s_{\mu\nu}\phi^J\Gamma^{\mu\nu}\epsilon + 4g^{Ir}b_{Jrs}Y^s_a\phi^J\rho^a \epsilon + 4d_{rs}^Ih_J^rh_K^s\phi^J\phi^K\epsilon = 0~.
\ee
Adding this to (\ref{int12}) gives
\be
\left[ D_\mu D^\mu \phi^I + \frac{1}{2}d_{rs}^I\mathcal{F}_{\mu\nu}^r\mathcal{F}^{s,\mu\nu} - 4d_{rs}^I Y^r_aY^{s,a} + 3d_{rs}^Ih_J^rh_K^s\phi^J\phi^K \right]\epsilon&&~\nn
+\frac{1}{2}\mathcal{F}^r_{\mu\nu}g^{Is}b_{Jsr}\phi^J\Gamma^{\mu\nu}\epsilon + 4g^{Ir}b_{Jrs}Y^s_a\phi^J\rho^a \epsilon &&~\nn
-\frac{1}{12}\left[ D_\mu \mathcal{H}^I_{\nu\rho\sigma} - \frac{3}{2}d_{rs}^I\mathcal{F}_{\mu\nu}^r\mathcal{F}_{\rho\sigma}^s\right] \epsilon^{\mu\nu\rho\sigma}{}_{\lambda\tau}\Gamma^{\lambda\tau}\epsilon&=&0~.
\label{int123}
\ee
The Bianchi identity for the 3-form field strength is given as
\be
D_{[\mu}\mathcal{H}_{\nu\rho\sigma]}^I &=& \frac{3}{2}d_{rs}^I\mathcal{F}_{[\mu\nu}^r\mathcal{F}_{\rho\sigma]}^s + \frac{1}{4}g^{Ir}\mathcal{H}_{\mu\nu\rho\sigma r}^{(4)}~,
\ee
using this (\ref{int123}) becomes
\be
\left[ D_\mu D^\mu \phi^I + \frac{1}{2}d_{rs}^I\mathcal{F}_{\mu\nu}^r\mathcal{F}^{s,\mu\nu} - 4d_{rs}^I Y^r_aY^{s,a} + 3d_{rs}^Ih_J^rh_K^s\phi^J\phi^K \right]\epsilon&&~\nn
+ 4g^{Ir}b_{Jrs}Y^s_a\phi^J\rho^a \epsilon \nn
+\frac{1}{2}\left[g^{Is}b_{Jsr}\mathcal{F}^r_{\mu\nu}\phi^J - \frac{1}{4!}\epsilon_{\mu\nu\rho\lambda\sigma\tau}g^{Ir}\mathcal{H}_r^{(4)\rho\lambda\sigma\tau}\right]\Gamma^{\mu\nu}\epsilon &=&0~.
\ee
The first line on the lhs gives the scalar field equation, the second line the $Y^r$ equations and the third line gives the relation between the 2-form and the 4-form field strengths.

\end{document}